\documentclass[useAMS,usenatbib]{mn2e}
\usepackage{graphicx,amsmath,multirow,amssymb}
\usepackage{natbib}
\newcommand{\comment}[1]{}

\def\simgt{\lower.5ex\hbox{$\; \buildrel > \over \sim \;$}}
\def\simlt{\lower.5ex\hbox{$\; \buildrel < \over \sim \;$}}

\title[Stardust from AGB and Super--AGB stars]{Dust formation around AGB and SAGB stars: a trend with
metallicity?}
\author[Ventura et al.]{P. Ventura,$^1$ M. Di Criscienzo,$^1$ R. Schneider,$^1$ 
R. Carini,$^{1,2}$ R. Valiante,$^1$ 
\newauthor
F. D'Antona,$^1$ S. Gallerani,$^1$ R. Maiolino,$^3$ A. Tornamb\'e$^1$  \\
$^1$INAF -- Osservatorio Astronomico di Roma, Via Frascati 33, 00040, Monte Porzio Catone (RM), Italy \\
$^{2}$Dipartimento di Fisica, Universit\`a di Roma ``La Sapienza'', P.le Aldo Moro 5, 00143, 
Roma, Italy \\
$^{3}$Cavendish Laboratory, University of Cambridge, 19 J.J. Thomson Ave., Cambridge CB3 OHE, UK}

\begin{document}

\date{Accepted, Received; in original form }

\pagerange{\pageref{firstpage}--\pageref{lastpage}} \pubyear{2012}

\maketitle

\label{firstpage}

\begin{abstract}
We calculate the dust formed around AGB and SAGB stars of metallicity Z=0.008 by following the 
evolution of models with masses in the range $1 M_{\odot}\leq M\leq 8 M_{\odot}$ through the thermal 
pulses phase, and assuming that dust forms via condensation of molecules within a wind expanding 
isotropically from the stellar surface.
We find that, because of the strong Hot Bottom Burning (HBB) experienced, high mass models produce 
silicates, whereas lower mass objects are predicted to be surrounded by carbonaceous grains; the 
transition between the two regimes occurs at a threshold mass of $3.5 M_{\odot}$. These findings 
are consistent with the results presented in a previous investigation, for Z=0.001. However, in the 
present higher metallicity case, the production of silicates in the more massive stars continues for 
the whole AGB phase, because the HBB experienced is softer at Z=0.008 than at Z=0.001, thus the oxygen 
in the envelope, essential for the formation of water molecules, is never consumed completely.
The total amount of dust formed for a given mass experiencing HBB increases with metallicity, 
because of the higher abundance of silicon, and the softer HBB, both factors favouring a higher 
rate of silicates production. This behaviour is not found in low mass stars, because the carbon 
enrichment of the stellar surface layers, due to repeated Third Drege Up episodes, is almost
independent of the metallicity.
Regarding cosmic dust enrichment by intermediate mass stars, we find that the cosmic yield at 
Z=0.008 is a factor $\sim$5 larger than at Z=0.001. In the lower metallicity case carbon dust 
dominates after $\sim$300~Myr, but at Z=0.008 the dust mass is dominated by silicates at all times, 
with a prompt enrichment occurring after $\sim$40~Myr, associated with the evolution of stars with 
masses $M\sim 7.5-8 M_{\odot}$. These conslusions are partly dependent on the assumptions concerning 
the two important macro--physics inputs needed to describe the AGB phase, and still unknowm from 
first principles: the treatment of convection, which determines the extent of the HBB experienced 
and of the Third Dredge--up following each thermal pulse, and mass loss, essential in fixing the 
time scale on which the stellar envelope is lost from the star.

\end{abstract}


\section{Introduction}
A reliable estimate of the nature and the amount of dust produced by stars of
intermediate mass ($1M_{\odot} \leq M \leq 8M_{\odot}$) proves essential for a number 
of scientific issues. These stars are believed to be the dominant stellar sources
of dust in the present-day Universe and, contrary to previous claims, their contribution
to dust enrichment can not be neglected even at redshift $z > 6$ \citep{valiante09, valiante11}.
While the formation of dust in the ejecta of core-collapse supernovae has received much
attention, both on the theoretical \citep{bianchi07, hirashita08, todini01} and observational 
side \citep{dunne09, morgan03, rho08}, models for the
nucleation of dust grains in the atmospheres of intermediate mass stars have been computed
either assuming synthetic stellar models (Ferrarotti \& Gail 2001, 2002, 2006) or 
exploring a single value for the initial stellar metallicity \citep{paperI}. 
In order to properly include their contribution in chemical evolution models with dust,
the mass and composition of dust grains released by each star as a function of its
mass and metallicity need to be known. In addition, the corresponding size distribution 
function allows to compute the extinction properties associated with these grains, which is 
fundamental information required for correctly interpreting the optical-near infrared 
properties of high-z quasar and gamma ray burst spectra 
\citep{gallerani10, maiolino04, stratta11}. 

Intermediate mass stars, after the end of the core--Helium burning phase, are 
nuclearly supported by a H--burning shell above the degenerate core, 
composed of carbon and oxygen \citep{iben75, iben76}. 
He--burning occurs periodically in the helium--rich buffer, in conditions of
thermal instability \citep{sh65, sh67}; this motivates the term ''Thermal Pulse'' (TP) 
used to describe these episodes. Due to the position in the HR diagram, this evolutionary
phase is known as Asymptotic Giant Branch (hereinafter AGB); AGB stars gradually
loose all their envelope, ending their evolution as White Dwarfs (WD). During the
AGB phase, characterized by strong mass loss, the conditions are most favourable 
for the condensation of gas molecules into dust.

The series of papers by the Heidelberg group \citep{gs99, fg01, fg02, fg06, zg08, zg09}
set the framework to describe the dust formation process in the environment
of AGBs. The scheme is based on a model of an expanding wind, whose thermodynamical
structure is determined by the physical parameters of the central object, i.e. surface gravity,
effective temperature, and the rate at which mass loss occurs. 

In the first paper of this series (Ventura et al. 2012, hereinafter paper I), we made
a step forward, by applying the wind modelling from \citet{fg06} to AGB models calculated
with a full integration of the whole stellar structure, more suitable than the synthetic
technique to deal with phenomena based on the thermodynamical coupling between the
internal, degenerate core, and the outer convective zone. The typical example is the
Hot Bottom Burning \citep{renzini81}, i.e. the series of proton--capture 
reactions at the bottom of the convective zone, once the temperature in that region reaches
$T_{\rm bce} \sim 40-50$~MK. The HBB phenomenon, active in models where the stellar
mass exceeds a threshold value of $\rm \sim 3-4~M_{\odot}$, is relevant for the production of
elements as well as the physical properties of the star. As regarding the surface chemistry, 
the main effect of HBB is the depletion of the surface carbon, and possibly 
(when $T_{\rm bce} > 80$~MK) oxygen, which prevents the formation of a C--star. The evolution 
of the star is also affected by HBB, because it is accompanied by a steep increase in
the luminosity and mass loss of the star \citep{blocker91}, so that the evolution
becomes faster, and only a limited number of Thermal Pulses (TPs) is experienced 
(see, e.g., \citet{vd09}).

A strong HBB naturally limits the effects of the alternative mechanism that can modify the
surface chemistry of AGBs, i.e. the Third Dredge--Up (TDU), which is the inward
penetration of the envelope in the phases following the TP, when the surface convection 
can reach layers previously exposed to $3\alpha$ nucleosynthesis \citep{iben75, lattanzio86, 
lattanzio89, wood81}. The main effect of the TDU is a great increase in the surface carbon, that 
eventually may become more abundant than oxygen, creating a C--star.

The kind of dust formed is extremely sensitive to which of the two afore mentioned mechanisms
dominate the evolution of the surface chemistry of the star: repeated TDU episodes favour
the formation of carbon--rich grains, whereas HBB destroys carbon, enabling the production of
silicates.

In paper I, we showed that under HBB conditions the formation of carbon--rich dust is inhibited, 
and only silicates are produced. This introduces a dichotomy in the type of dust
produced by AGBs: the more massive stars produce silicates, whereas lower-mass objects
produce carbon dust. The analysis 
outlined interesting differences with the results by \citet{fg06}, that can
all be understood on the basis of the different HBB experienced by the stars.

In this paper, we compare the results of paper I, that were limited to a 
single stellar metallicity ($Z=0.001$), to AGB and Super-AGB (SAGB, stars of higher masses
that evolve on a core made of Oxygen and Neon) stellar models with initial metallicity
$Z=0.008$. The main goal is to understand how the dust formed around AGBs and SAGBs
changes with the initial chemical composition of the stars, and whether the basic difference between
the type of dust formed around massive AGBs and lower--mass stars (producing, respectively,
silicates and carbon--rich dust) persists. To this aim, we calculated a new set of AGB and
SAGB models of metallicity $Z=0.008$, and applied the same scheme used by \citet{fg06}
and in paper I to calculate the dust produced in their surroundings.

The paper is organized as follows. Section 2 describes the modelling of the stellar evolution 
and of the structure of the wind; the evolution properties of AGB and SAGB stars are 
discussed in Section 3; the results concerning the quantity and the type of dust
produced are given in Section 4, whereas Section 5 deals with the uncertainties 
associated to the choices of the input macro--physics and of the optical constants of the 
silicates; finally, the results are discussed and commented in Section 6.

\section{The model}
\label{sec:refmodel}
Dust grains are assumed to form from condensation of gas molecules present in the
expanding winds. The description of this process, and the determination of the kind 
and quantity of dust species formed, requires an accurate modelling of the evolution 
of the star along the AGB (or SAGB) phase, from which we obtain the temporal variation 
of mass, mass loss rate, luminosity, effective temperature, and the surface chemical composition:
all this information is used to model the thermodynamical and chemical structure
of the wind, hence to describe the dust formation process. 

A word of caution in needed here. Mass loss during the AGB phase is here obtained by means 
of semi--empirical descriptions, that include some parameters, calibrated based on the 
comparison with the observations. More precisely, we assume a mass-loss rate and then 
compute the corresponding wind properties and the required degree of dust condensation. 
In reality, however, the mass loss is believed to be a consequence of dust formation, which 
makes such modeling approach seem somewhat circular. But we like to stress that this is the 
best one can do at present at a reasonable computational cost.

In what follows we provide a brief description of the numerical scheme adopted; we refer to 
\citet{ventura98} and to paper I for further details.

\subsection{Stellar evolution modelling}
The stellar models presented in this work were calculated by means of the ATON 
stellar evolution code, in the version described in \citet{ventura98}.

The extension of the convective regions was determined via the classic Schwartzschild criterium,
stating that the convective instability is favoured by the condition $\nabla_{\rm rad} > 
\nabla_{\rm ad }$, where $\nabla$ is the logarithmic gradient of
temperature with respect to pressure.
The temperature gradient within regions unstable to convective motions was determined
by means of the Full Spectrum of Turbulence (hereinafter FST) model for turbulent 
convection \citep{cm91}. In the convective regions
where the temperature is sufficiently large to allow the ignition of nuclear reactions,
we follow the variation of the chemistry by coupling the convective motions with the
process of nuclear burning, by means of a diffusive approach, according to the scheme by
\citet{cloutman}. The velocities determined via the convection modelling are used to
find the diffusion coefficients, and also to provide the extension of a possible
extra--mixing zone, based on an exponential decay of velocities from the formal
convective/radiative interface, determined on the basis of the Schwartzschild criterium: the
scale for the decay of velocities is $l=\zeta H_p$, where $\zeta$ is the free parameter
associated to the extension of the extra--mixed region. We assumed
$\zeta=0.02$ during the evolution before the beginning of the Thermal Pulses phase, 
in agreement with the calibration given in \citet{ventura98}. For what concerns the 
AGB phase, we compare the results found ignoring any overshoot from the convective
borders, with those obtained by assuming a tiny extra--mixing from the convective shells 
developed as a consequence of the ignition of the Thermal Pulse, with $\zeta=0.001$.

Mass loss was modelled following \citet{blocker95}, with the free parameter entering this
recipe set to $\eta_R=0.02$. The value for $\eta_R$ was calibrated specifically
for this range of masses and for this metallicity via a comparison between the observed 
and the predicted luminosity function of lithium rich sources and of carbon stars in the
Large Magellanic Cloud \citep{ventura99, vdm2000}.

The OPAL radiative opacities \citep{opal} were used for temperatures above $10^4$~K,
whereas for smaller temperatures we used the AESOPUS tool described in \citet{marigo09}.
This choice allows to account for the increase in the opacity associated with the
enrichment in the surface carbon, as a consequence of the TDU: this is particularly
important in the description of the physical properties of low--mass AGBs, as described
in details by \citet{vm09} and \citet{vm10}. The interested reader may find in paper I
a detailed discussion on how the modelling of the absorption coefficients in C--rich
mixtures affects dust production by AGBs that reach the C--star stage.

We follow the nucleosynthesis evolution of 30 elements, from hydrogen to alluminum,
with the most relevant isotopes entering the p-p and $3\alpha$ chains, and the CNO cycle.
The nuclear cross sections are taken from the NACRE compilation \citep{angulo},
with a few exceptions, the most relevant for this study being the rate of the proton
capture reaction by nitrogen nuclei, taken from \citet{formicola}. 

The models presented in this work were calculated with an overall metallicity
$Z=0.008$, a helium content $Y=0.26$, and a mixture scaled according to \citet{gs98},
with an $\alpha-$enhancement $\rm [\alpha/Fe]=+0.2$.

\subsection{Dust production and the stellar wind}
The scheme adopted to model the structure of the wind at a given phase during the AGB
evolution is the same as in paper I, based on the description given in the series
of papers by \citet{fg01, fg02, fg06}.

The wind is modelled via two differential equations describing the radial behaviour of
the velocity of the gas and the optical depth. Indicating the mass, luminosity, effective 
temperature, radius, mass loss rate of the star with $M$, $L$, $T_{eff}$, $R_*$, and $\dot M$, 
we have:

\begin{equation}
v{dv\over dr}=-{GM\over r^2}(1-\Gamma),
\label{eqvel}
\end{equation}

\begin{equation}
{d \tau \over dr}=-\rho k {R_*^2\over r^2},
\label{eqtau}
\end{equation}
\noindent
where

\begin{equation}
\Gamma={kL\over 4\pi cGM}.
\label{eqgamma}
\end{equation}
\noindent

Equations \ref{eqvel} and \ref{eqtau} are completed by the law of mass conservation:

\begin{equation}
\dot M=4\pi r^2 \rho v,
\label{eqmloss}
\end{equation}
\noindent
and by the radial variation of temperature in the approximation of a spherically simmetric, 
grey wind \citep{lucy76}:

\begin{equation}
T^4={1\over 2}T_{\rm eff}^4 \left[ 1-\sqrt{1-{R_*^2\over r^2}}+{3\over 2}\tau \right],
\end{equation}
\noindent
where $k$ is the flux--averaged extinction coefficient of the gas--dust mixture, and can be 
expressed as,

\begin{equation}
k=k_{\rm gas}+\sum_i f_ik_i,
\label{eqk}
\end{equation}
\noindent
and $k_{\rm gas}=10^{-8}\rho^{2/3}T^3$ \citep{bell94}. The sum in eq.~(\ref{eqk}) is 
extended to all the dust species considered: the $f_i$ terms give the degrees of condensation 
of the key--elements for each dust species, whereas $k_i$ represent their corresponding 
extinction coefficients. 
This simple opacity law is used in the wind model since the AESOPUS tool does not give 
opacities for temperatures below 2000K. This is not an entirely correct use of the 
\citet{bell94} formula, which of course adds physical inconsistency to the model. But since the gas 
opacity 
is typically much smaller than the overall dust opacity, the effects are small. In fact, it would 
have made little difference if we had assumed $k_{\rm gas}=0$ instead, as we integrate from the 
condensation radius (where dust formation region starts and dust opacity dominates) out to a 
distance $10^4R_*$ away from the star (where the gas opacity is very low).

We consider the thermal gas pressure force to be negligible beyond the condensation radius, 
compared to the radiative acceleration. Hence, we omit the pressure term in the equation 
of motion, but assume that there is a nonzero flow over the inner boundary (we assume 
$v_0 = 2$km/s, which is similar to the expected sound speed) despite the fact that there 
is formally no gas pressure at the inner boundary. Obviously, this is physically 
inconsistent, but the effect on the results is usually small and assuming a nonzero flow 
speed at the inner boundary simplifies the numerical solution of equations (1-6) 
considerably. We note, however, that this assumption may be somewhat problematic for 
critical winds ($\Gamma \sim 1$), which are slow and may have outflow speeds which are 
comparable to the assumed sound speed at the inner boundary.

Dust growth takes place by vapour deposition on the surface of some 
pre-formed seed nuclei, assumed to be nano--meter sized spheres. 
The precise chemical nature of these seed grains is not important in what 
follows. We consider various types of dust, depending on the surface chemical composition of 
the star. In oxygen--rich winds, we consider olivine, pyroxene, quartz and iron grains, whereas 
for C--rich environments we account for the presence of solid carbon, silicon carbide and 
iron grains. For each condensate, we define a key element whose abundance is the minimum among all 
the elements necessary to form the corresponding dust aggregate. Silicon is the key element for 
olivine, pyroxene, quartz and silicon carbide, whereas iron and carbon are the key elements for 
iron dust and solid carbon. All the species considered, with the corresponding condensation 
reactions and key--elements, are listed in Table~1. 

\begin{figure}
\resizebox{1.\hsize}{!}{\includegraphics{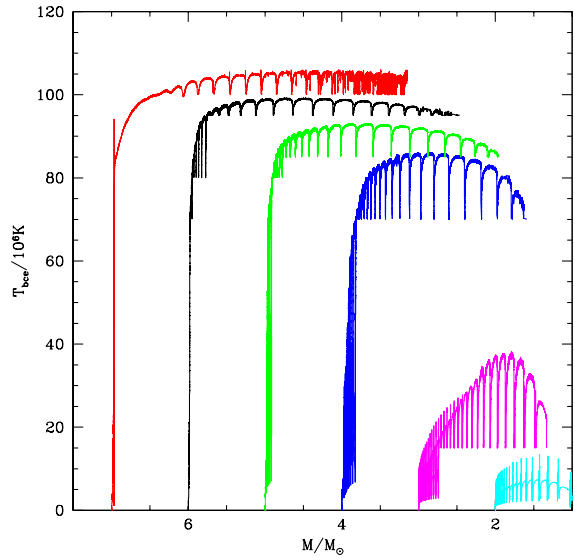}}
\vskip-40pt
\caption{Variation during the Thermal pulses phase of the temperature at the bottom
of the external convective zone as a function of the total stellar mass of models with 
initial mass $\rm 7~M_{\odot}$ (red), $\rm 6~M_{\odot}$ (black), $\rm 5~M_{\odot}$ (green), 
$\rm 4~M_{\odot}$ (blue), $\rm 3~M_{\odot}$ (magenta), $\rm 2~M_{\odot}$ (cyan). Note that models
experiencing HBB keep the same temperature for the vast majority of the time during
which they undergo the strongest mass loss.
}
\label{ftbce}
\end{figure}

The growth rate of each dust species depends on the competition between the formation
and destruction rates. The former is evaluated on the basis of the number density of the
key--element and the thermal velocity of the corresponding molecule, whereas the latter
is calculated via the difference between the formation enthalpy of the dust species and of
the individual molecules concurring to the formation process. All the references concerning
the thermodynamic quantities considered can be found in paper I.

\begin{table*}
\begin{center}
\caption{Dust species considered in the present analysis, their formation reaction and
the corresponding key--elements (see text).} 
\begin{tabular}{l|l|l|l}
\hline
\hline 
Grain Species & Formation Reaction & Key--element \\
\hline
Olivine & 2$x$Mg +2(1-$x$)Fe+SiO+3H$_2$O $\rightarrow$ Mg$_{2x}$Fe$_{2(1-x)}$SiO$_4$ + 3H$_2$ & Si \\ 
Pyroxene & $x$Mg +(1-$x$)Fe+SiO+2H$_2$O  $\rightarrow$ Mg$_{x}$Fe$_{(1-x)}$SiO$_3$ + 2H$_2$ & Si \\
Quartz & SiO + H$_2$O $\rightarrow$ SiO$_2(s)$ +H$_2$ & Si \\  
Silicon Carbide & 2Si + C$_2$H$_2$ $\rightarrow$ 2 SiC + H$_2$ & Si \\ 
Carbon & C $\rightarrow$ C$(s)$ & C \\
Iron & Fe $\rightarrow$ Fe$(s)$ & Fe \\
\hline 
\hline 
\end{tabular}
\end{center}
\label{tabrates}
\end{table*}

The regions close to the central star are generally too hot to allow for dust condensation,
the destruction rate exceeding by far the production rate. At a distance of 3-4 stellar radii
from the surface of the star, the temperature drops below 1000~K, and dust formation
occurs. The consequent increase in the opacity accelerates the wind via the strong radiation 
pressure, and halts dust formation. We therefore expect the sizes of the different grain species 
and the terminal velocity of the wind to reach an asymptotic behaviour.

Because the assumed initial dimension of the grains, $a_0$, is much smaller than the
size reached by the various grains in the asymptotic regime, the results are practically
independent of the choice for $a_0$: simulations based on different values for $a_0$ 
(still in the nano-sized regime) lead to the same result.

A further initial condition, imposed by the lack of a nucleation theory to be applied for
any of the dust species considered, is the initial density of seed grains, $n_d$. In
agreement with paper I, we assume $n_d=3\times 10^{-13}n_H$ (note that in Section 2.2.2
of paper I, due to a typo, it is given $n_d=3\times 10^{-4}n_H$ instead), which, as order of
magnitude, reflects the typical number densities of grains in the outflows of AGBs
\citep{knapp85}. A higher $n_d$, for a given size of the grains, corresponds to a higher
degree of condensation of the key--species into dust, and consequently to a smaller
density of the various molecules in the wind (see \citet{fg06}, Eq.20--30). This is not
relevant as far as the degree of condensation is small, and has the largest impact in
those cases where dust condensation is most efficient: i.e. the SAGB models presented here,
which we will see is the most efficient producers of dust. A change in $n_d$ by a factor
2 determines a variation in the grain size of the order of $\sim 10\%$.

To summarize, the amount of dust produced and its composition are mainly determined by the following 
quantities:
\begin{enumerate}
\item{The physical parameters of the central star and, in particular, the luminosity, effective
temperature and the mass loss rate; these determine the radial variation of the thermodynamics of the wind.}

\item{The surface chemistry of the star, that is relevant in the determination of the
dominant dust species (either silicates or carbon dust, according to the C/O ratio),
and in the quantity of dust formed (via the mass fractions of the key--elements).}

\item{The description of the absorption and scattering processes for the various elements,
that determine the extinction coefficient.}
\end{enumerate}

\section{The evolutionary properties of AGB and SAGB models with $Z=0.008$}
The main physical features of the evolution of AGB stars are discussed in the reviews by 
\citet{karakas11}, \citet{herwig05}; further details on the efficiency of the TDU and
the achievement of the C--star stage can be found in \citet{stancliffe04}, \citet{stancliffe05},
\citet{karakas10} (and references therein). An exhaustive description of the SAGB phase can be 
found in the classic paper by \citet{garciaberro97}, and the more recent investigations by
\citet{gilpons07}, \citet{siess07, siess10}.

\begin{figure*}
\begin{minipage}{0.45\textwidth}
\resizebox{1.\hsize}{!}{\includegraphics{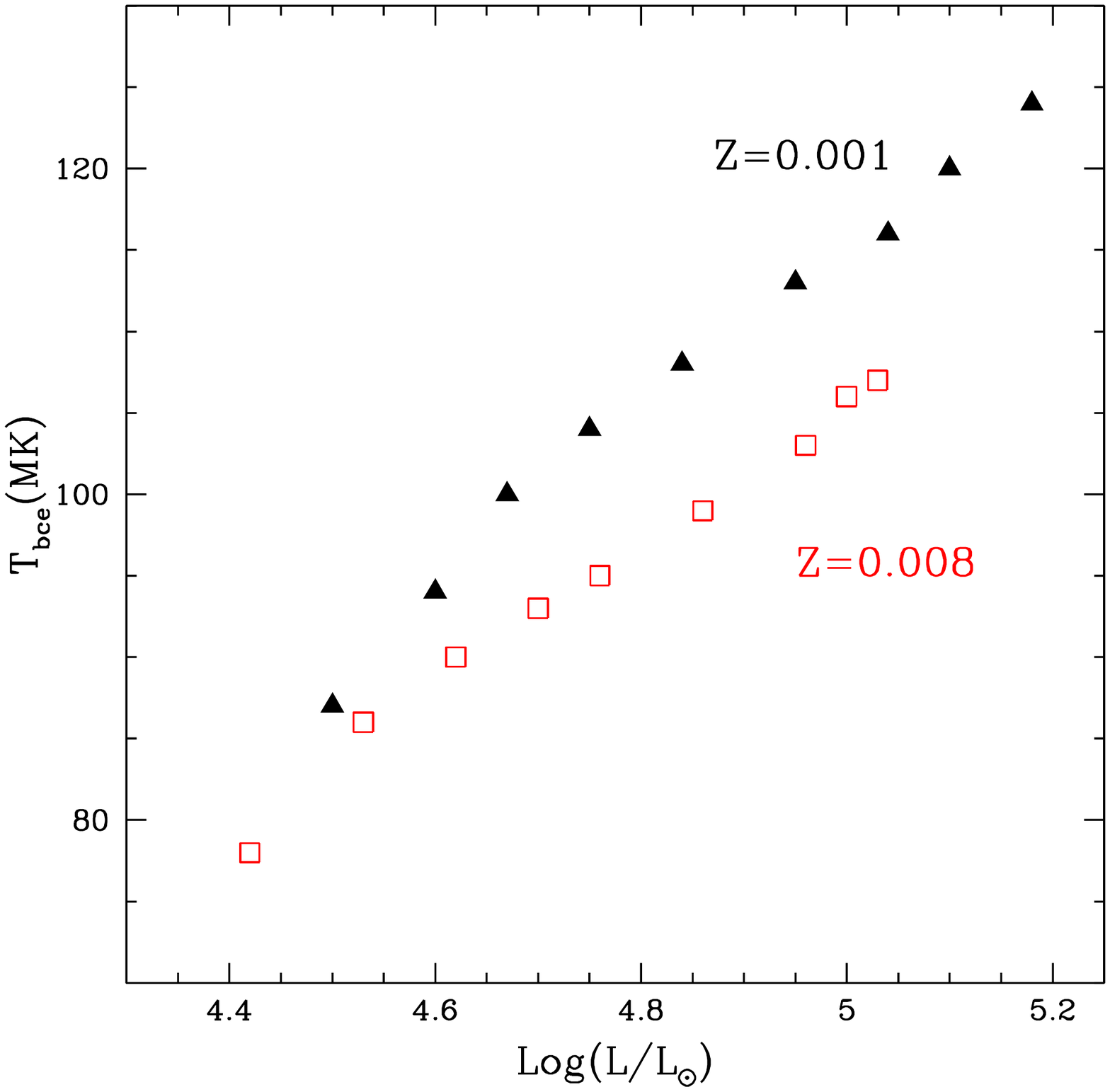}}
\end{minipage}
\begin{minipage}{0.45\textwidth}
\resizebox{1.\hsize}{!}{\includegraphics{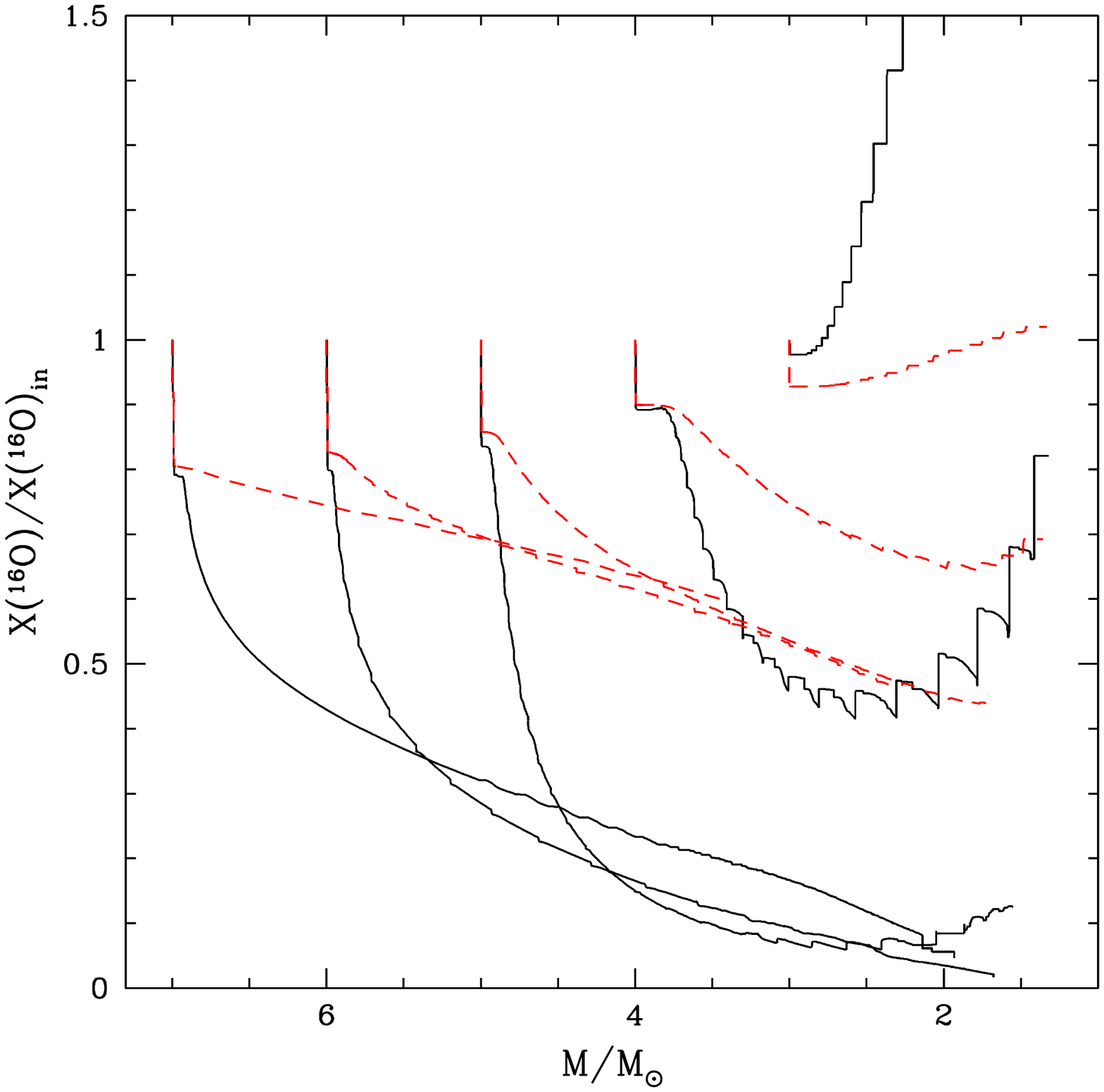}}
\end{minipage}
\vskip-40pt
\caption{{\it Left panel}: the average temperature at the bottom of the convective envelope as a
function of the peak luminosity reached during the Thermal Pulses phase in AGB and
SAGB models of metallicity $Z=0.008$ studied in the present investigation (open squares)
and those presented in \citet{vd09, vd11} (red,full triangles). {\it Right panel}: surface oxygen
mass fraction (divided by the initial abundance) as a function of the stellar mass
for the models with $Z=0.001$ discussed in paper I (solid, black), and the models presented
here (dashed, red).
}
\label{fagb}
\end{figure*}

Here we provide a brief description of the main features of the evolution of
the AGB and SAGB stars with initial metallicity $Z=0.008$, 
that prove essential in the understanding of the dust formation process.

The threshold mass separating the AGB and SAGB regimes is $\rm 6.5 M_{\odot}$. Less
massive models evolve on a CO core, whereas higher masses undergo off--center carbon ignition, 
and eventually develop a degenerate core, made up of oxygen and neon. This finding is partly 
dependent on the amount of overshoot from the convective core: when the extra--mixing 
is neglected, the above limit shifts upwards by $\rm \sim 1 M_{\odot}$.

\citet{vd05} outlined the key role played by the treatment of the convection in
the extent of HBB, i.e. the nuclear activity that develops at the bottom of the convective
envelope once the temperature exceeds $30-40$~MK. The FST modelling of convection, used in the
present work, leads easily to HBB conditions for models whose initial mass exceeds a
threshold value \citep{dm96, vd05} that depends on the metallicity of the star and on the 
assumptions concerning the overshoot from the border of the convective core during the two
major phases of hydrogen and helium burning. The relevance of HBB in this context is twofold: 
a) even in moderate HBB conditions carbon is destroyed in the envelope of the star, 
thus leaving the only possibility for the formation of
silicate--type dust; b) HBB is accompanied by an increase in the luminosity of the star, 
which, given the steep dependence on luminosity of the Bl\"ocker's mass loss description 
($\dot M \propto M^{3.7}$), favours a higher rate of mass loss; the relationship 
between $\dot M$ and density of the gas in the wind (see Eq.4), will make dust formation seem
more efficient\footnote{Note that this result is a consequence of the scheme we follow, i.e. 
adopting an empirically determined mass loss treatment, and finding out the dust formed accordingly.
As stated in Sect.2, the correct treatment would be to determine the mass loss rate on the basis
of the dust formed}.

The variation during the AGB (or SAGB) phase of the temperature at the bottom of the
convective envelope of models differing in their initial mass is shown in Fig.~\ref{ftbce}.
The choice of the mass of the star as abscissa allows to identify the physical conditions 
at the time when most of the mass loss occurs. 
We can clearly identify a threshold mass of $\rm 3.5~M_{\odot}$ separating the more massive 
objects, experiencing HBB, from their lower mass counterparts, 
whose temperature at the bottom of the convective envelope remains below $40$~MK. Interestingly,
we note that models experiencing HBB maintain an approximately constant temperature during
most of the time when they loose their mass: this allows us to understand the nucleosynthesis
experienced, and the corresponding variation of the surface chemistry.

To understand the effects of the initial metallicity of the stars on the HBB phenomenology, 
in the left panel of Fig.~\ref{fagb} we show, for each star, the temperature at the bottom 
of the surface convective zone and the peak luminosity during the AGB evolution, which, in turn, 
is related to the core mass of the star. Models discussed in the present work are indicated as 
open squares, whereas the $Z=0.001$ models analyzed in paper I are represented by full triangles.

Lower $Z$ models evolve, for a given core mass (luminosity) at higher temperatures, thus 
allowing a stronger nucleosynthesis at the bottom of the convective envelope. 
This difference is confirmed in the right panel of Fig.~\ref{fagb}, showing the evolution of the 
surface oxygen with mass for models of different initial mass and metallicity 
$Z=0.008$ and $Z=0.001$. To illustrate the extent of oxygen depletion, we report
in the ordinate the ratio of the surface oxygen mass fraction to its initial value. All
the $Z=0.008$ models experiencing HBB ($\rm M > 3.5 M_{\odot}$) achieve only a modest depletion 
of the surface oxygen, in comparison to their lower $Z$ counterparts which, in some cases, 
reduce the oxygen in their envelope by a factor $\sim 10$. 
For both metallicities, the strongest depletion is reached for masses still in the AGB regime; 
in SAGBs mass loss rates and oxygen depletion proceed with comparable timescales: these stars loose 
a high percentage of their mass before a strong depletion of the surface oxygen is achieved
\citep{vd11}.

\begin{figure*}
\begin{minipage}{0.45\textwidth}
\resizebox{1.\hsize}{!}{\includegraphics{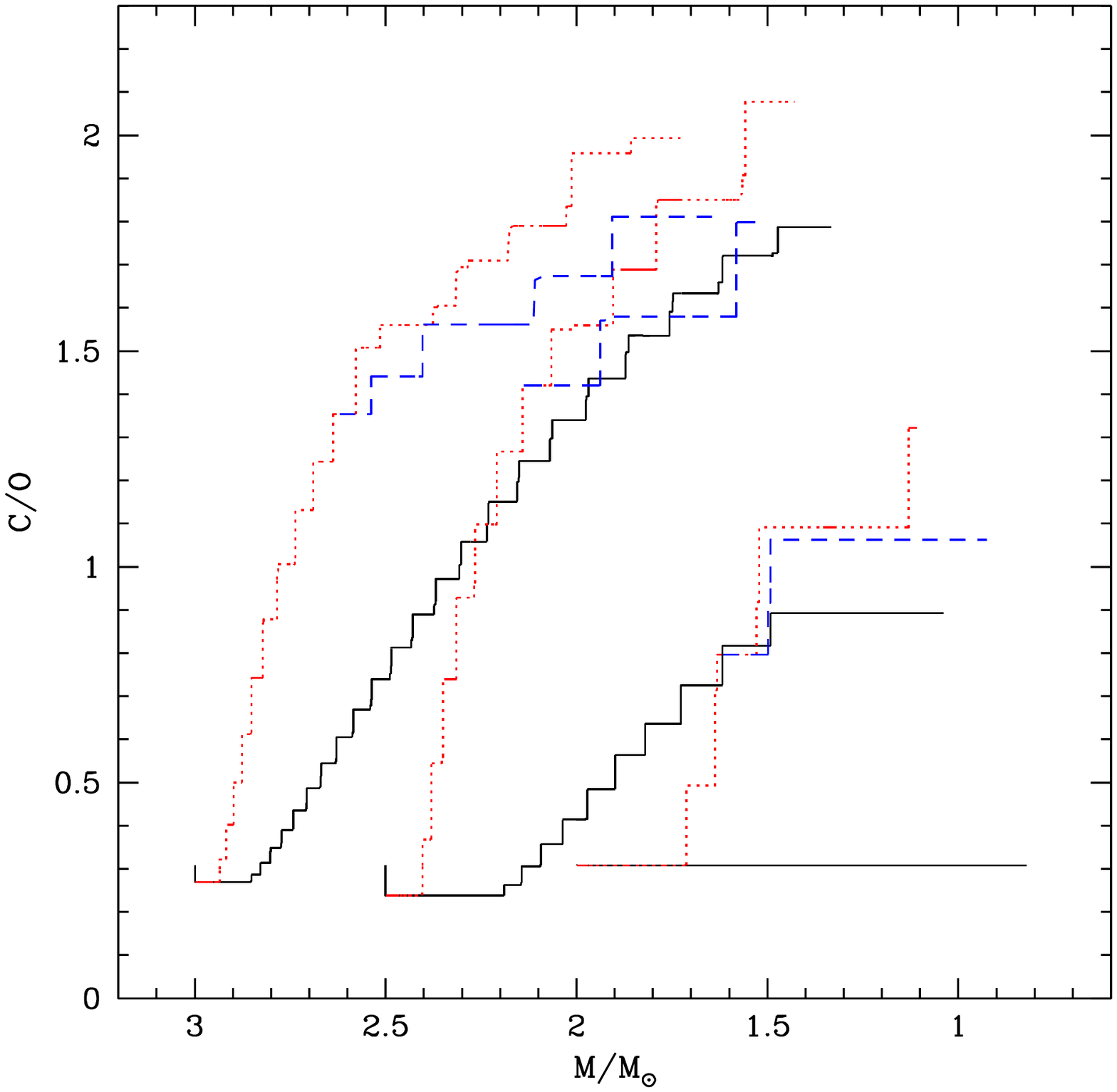}}
\end{minipage}
\begin{minipage}{0.45\textwidth}
\resizebox{1.\hsize}{!}{\includegraphics{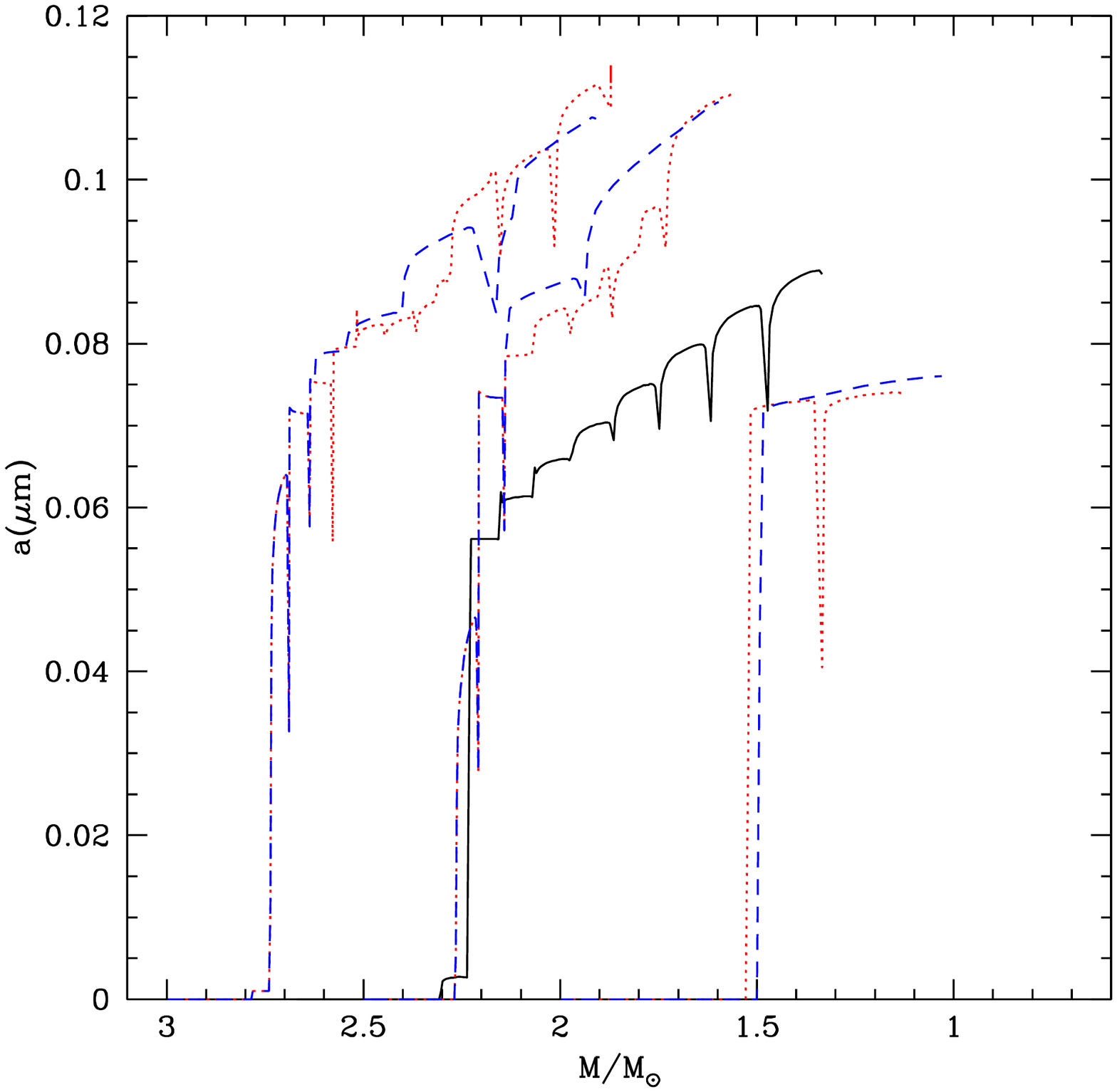}}
\end{minipage}
\vskip-40pt
\caption{Left: The variation of the C/O ratio in $Z=0.008$ models not experiencing HBB, with
initial masses 2,2.5, and 3M$_{\odot}$. Lower mass models are not shown, since they do
not experience any TDU in any case. The solid lines (black) indicate the results obtained when the borders 
of the convective region developed during each Thermal Pulse is fixed via the Schwarzschild criterium; 
dotted (red) lines indicate the evolution when a small extra-mixing is assumed (see text for details).
Dashed, blue lines indicate the variation of the C/O ratio when extra--mixing is adopted, and 
mass loss during the C--star stage is modeled with the treatment by \citet{watcher08}. Right: The
variation of thi size of carbon grains in the same models presented in the left panel.}
\label{fovsh}
\end{figure*}

In models with M$\leq 3$M$_{\odot}$, not experiencing any HBB, we expect only a
poor, if any, production of silicates, because the mass loss rates are extremely small, thus
preventing the possibility that a radiation pressure driven wind develops. 

\begin{figure*}
\begin{minipage}{0.45\textwidth}
\resizebox{1.\hsize}{!}{\includegraphics{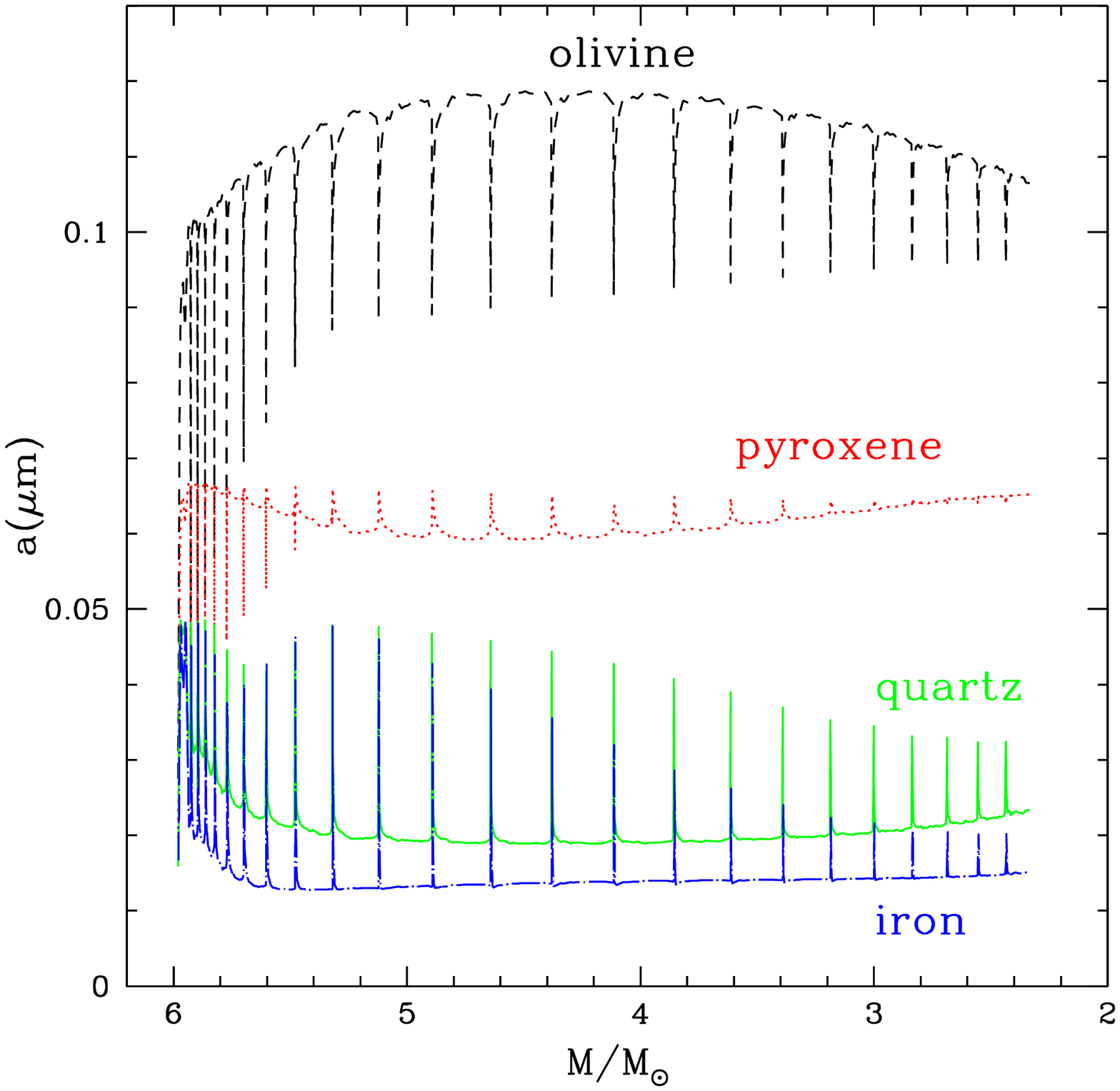}}
\end{minipage}
\begin{minipage}{0.45\textwidth}
\resizebox{1.\hsize}{!}{\includegraphics{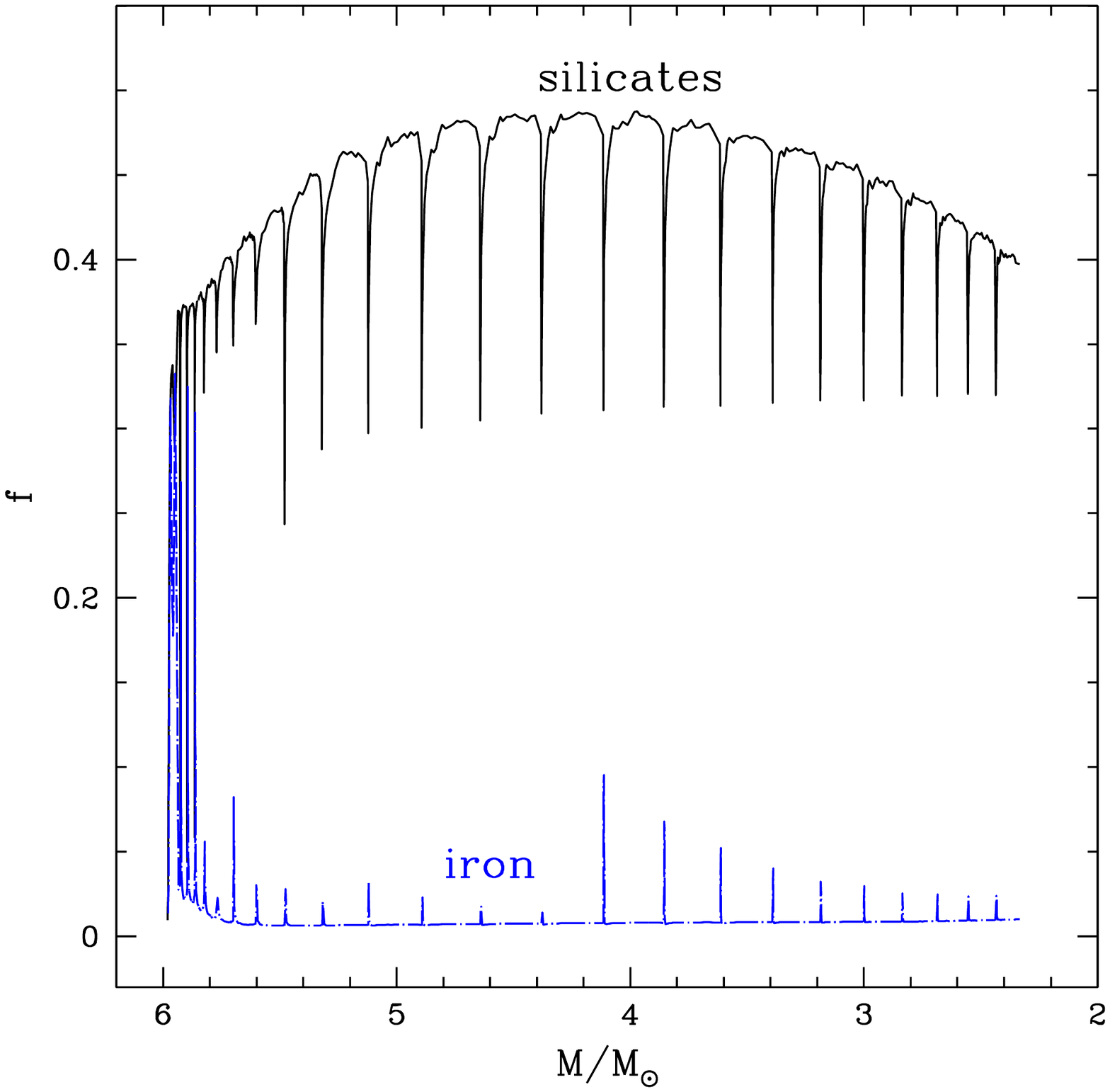}}
\end{minipage}
\vskip-40pt
\caption{{\it Left panel}: variation of the size of different grain species formed during the AGB 
evolution of a model with initial mass $\rm 6 M_{\odot}$: olivine (dashed, black), pyroxene (dotted, red),
quartz (solid, green), and iron (dot--dashed, blue).
{\it Right panel}: the fraction of silicon condensed into silicates (solid, black), and the fraction 
of iron condensed into iron grains (dot--dashed, blue) for the same $\rm 6 M_{\odot}$ stellar model.}
\label{f6msun}
\end{figure*}

The possibility for these stars to form dust is related to the TDU. A highly efficient 
TDU drives the bottom of the envelope to regions previously enriched by $3\alpha$ 
nucleosynthesis; the consequent increase in the carbon content in the envelope eventually 
leads to the C--star stage: this would eliminate the oxygen needed to form silicates, 
and favour the formation of carbon dust\footnote{When C/O$>1$, all the oxygen 
available is locked into CO molecules, which are highly stable, due to their large 
dissociation energy.}. 

The issue of forming carbon stars is a long-standing 
problem for stellar evolution theories. On the observational side, the 
investigations by \citet{groenewegen93}, \citet{marigo99}, \citet{izzard04},
indicated the core mass at which TDU must begin, and the efficiency of the inwards 
penetration of the convective mantle required to reproduce the luminosity function of 
carbon stars in the Magellanic Clouds.
Most of the evolution codes fail to achieve this result, because the TDU begins
late during the AGB evolution, with an efficiency smaller than required
by the observations \citep{mowlavi99, herwig00}. Due to the extreme sensitivity
of the results to the details of the numerical treatment of the convective
borders (e.g. the modality with which the convective/radiative interface is
determined, the spatial zoning near the convective
boundaries), the results obtained by various research groups are
substantially different \citep{straniero97}. An important step forward 
was made by \citet{stancliffe05}, that using a fully implicit and simultaneous
solution of the equations of stellar structure, nuclear burning, and diffusive
mixing, could reproduce the luminosity function of carbon stars in the LMC, with no 
need of any free parameter.

The models produced by our code, where the boundaries of convective regions is
determined via a straight application of the Schwartzschild criterium 
(i.e. $\zeta=0$), share the difficulty of obtaining a TDU with the required 
efficiency. As can be seen in the left panel of Fig.3 (solid lines), that shows the 
evolution of the surface C/O ratio,
the C-star stage is reached only by models with initial mass M=2.5,3M$_{\odot}$, in 
a late phase of the AGB evolution. On the other hand, to confirm the 
importance of the treatment of convective boundaries, we have shown that even
a modest extra-mixing from the convective shell formed during the TP leads
to very different results: the results indicated with dotted tracks in Fig.3
were obtained with $\zeta=0.001$, much smaller than the values invoked to
describe overshoot from the borders of the convective cores during the main
sequence phase (i.e. $\zeta=0.02$). The impact of the assumed extra--mixing can be understood 
by the comparison with the solid lines. The choice $\zeta=0.001$ favours the achievement of the 
C--star stage for the stars with initial mass between 2M$_{\odot}$ and 3M$_{\odot}$, 
reported in Fig.~\ref{fovsh}. We note that even a tiny overshoot from the shell developed
during each TP is sufficient to induce a deep TDU, which leads to a C/O ratio exceeding 
unity (the essential condition to produce solid carbon and silicon carbide) after
a few TPs, when only a $\sim 5-10\%$ of the stellar mass is lost. With the present choice
concerning $\zeta$, we find that stars with mass below 1.5M$_{\odot}$ never become
carbon stars.

A detailed tuning of the amount of extra-mixing needed to account for the observations is 
beyond the scope of the present investigations; we limit here to stress the sensitivity 
of the results to this choice, and we will describe how the dust formation process also 
depends on $\zeta$.

\section{Dust production}
The results discussed in the previous section indicate that one of the main findings of paper I
holds also at higher metallicites: stars whose initial mass is above the threshold value to allow HBB
conditions produce silicates, whereas lower mass objects produce carbon--rich dust.

We show in the left panels of Fig.~\ref{f6msun} the size of the dust grains 
formed around a typical model experiencing HBB, with an initial mass $\rm M = 6 M_{\odot}$. 
Olivine is the dust species most easily formed, followed by pyroxene, and by 
small quantities of quartz and iron. The spread between the size of the olivine grains
and of the other elements gets larger as the star evolves on the AGB. This is because
the increase in the mass loss rate is associated with a more efficient production of 
olivine; this, in turn, determines a faster acceleration of the wind, that prevents 
the formation of the other species. Note the periodic drop in the size of the olivine 
grains, when TPs occur, and the star looses mass more slowly.

In the right panel of Fig.~\ref{f6msun} we show, for the same $\rm 6 M_{\odot}$ stellar model, 
the fractions of silicon and iron condensed into dust. In the present set of $Z=0.008$ models the 
production of silicates continues for the whole AGB evolution; this is different compared 
to paper I, where it is shown that in massive AGBs silicates production stops (see paper 
I, left panel of Fig.~3). This difference can be understood from the right 
panel of Fig.~\ref{fagb}, showing 
that surface oxygen destruction is much softer in the present models than at lower $Z$. 
Unlike the $Z=0.001$ stars, here we never enter the situation in  which water, one of the key 
ingredients to form the silicates (see Table~\ref{tabrates}), is consumed, and therefore
the production of the silicates never stops.

\begin{figure}
\resizebox{1.\hsize}{!}{\includegraphics{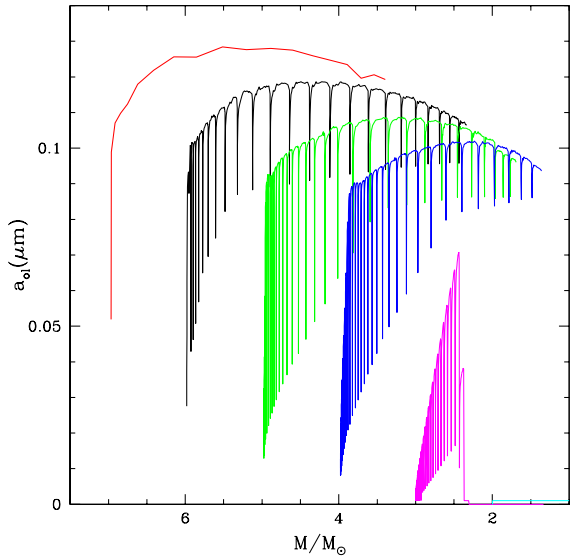}}
\vskip-40pt
\caption{Variation during the AGB (or SAGB) evolution of the size of olivine grains
formed around stars with different initial mass that experience HBB. The color code is 
the same of Fig.~\ref{ftbce}.
}
\label{foliv}
\end{figure}

Since olivine is the most abundant species, we may understand the trend with stellar mass
of the amount of silicates formed by comparing the size of the olivine grains in the
surroundings of AGB stars of different initial mass. This is shown in Fig.~\ref{foliv}. 

Higher mass models evolve at large luminosities. Given the steep slope of the
$\dot M(L)$ relationship provided by the Bl\"ocker's description, mass loss will
proceed faster. Because we keep the velocity
constant until the point where dust formation becomes possible (i.e. where for
at least one of the species considered the production term exceeds destruction), 
we find via Eq.4 a larger gas density there. This favours dust formation by increasing 
the number density of all the molecules entering the various sublimation reactions.
More massive models are therefore expected to produce grains with
larger size, as confirmed by the trend shown in Fig.~\ref{foliv}. The fraction 
f$_{sil}$ of silicon condensed into dust ranges from 35$\%$ at M=$3.5M_\odot$up to 50$\%$ 
in the SAGB regime. Because a great production of silicates affects iron condensation,
the trend with mass of the fraction of iron condensed is opposite: f$_{ir}$ varies
from 10$\%$ at low M, down to $\sim 2\%$ for SAGBs. 

The grain sizes obtained 
here for olivine (see figures 4 and 5) are marginally consistent with the grain sizes required to 
drive a wind according to H\"ofner (2008), but we note also that the grain sizes obtained at lower 
metallicity (paper I) are significantly smaller, which indicate some tension between our results 
and the grain-size requirement derived by H\"ofner (2008). The predictions by H\"ofner (2008)
have very recently been confirmed observationally by \citet{norris2012}, which also seems
to rule out any significant production of iron--rich silicates. However, the latter may
in part explain the discrepancy with our results, since iron--free silicates have
clearly different optical properties (lower opacity) that may favour growth to large
sizes as they are not heated by radiation as much as iron--rich silicates.

The larger depletion of surface oxygen in more massive stars has a negligible role here, 
because the abundance of the key element for silicates, i.e. silicon, remains almost constant
in all cases\footnote{The possibility that surface silicon increases in the AGB models experiencing 
the strongest HBB is discussed in \citet{vcd11}, who however find a modest (if any) increase 
in the silicon mass fraction, that would have only a negligible impact on the results presented 
here.}.

\begin{figure*}
\begin{minipage}{0.45\textwidth}
\resizebox{1.\hsize}{!}{\includegraphics{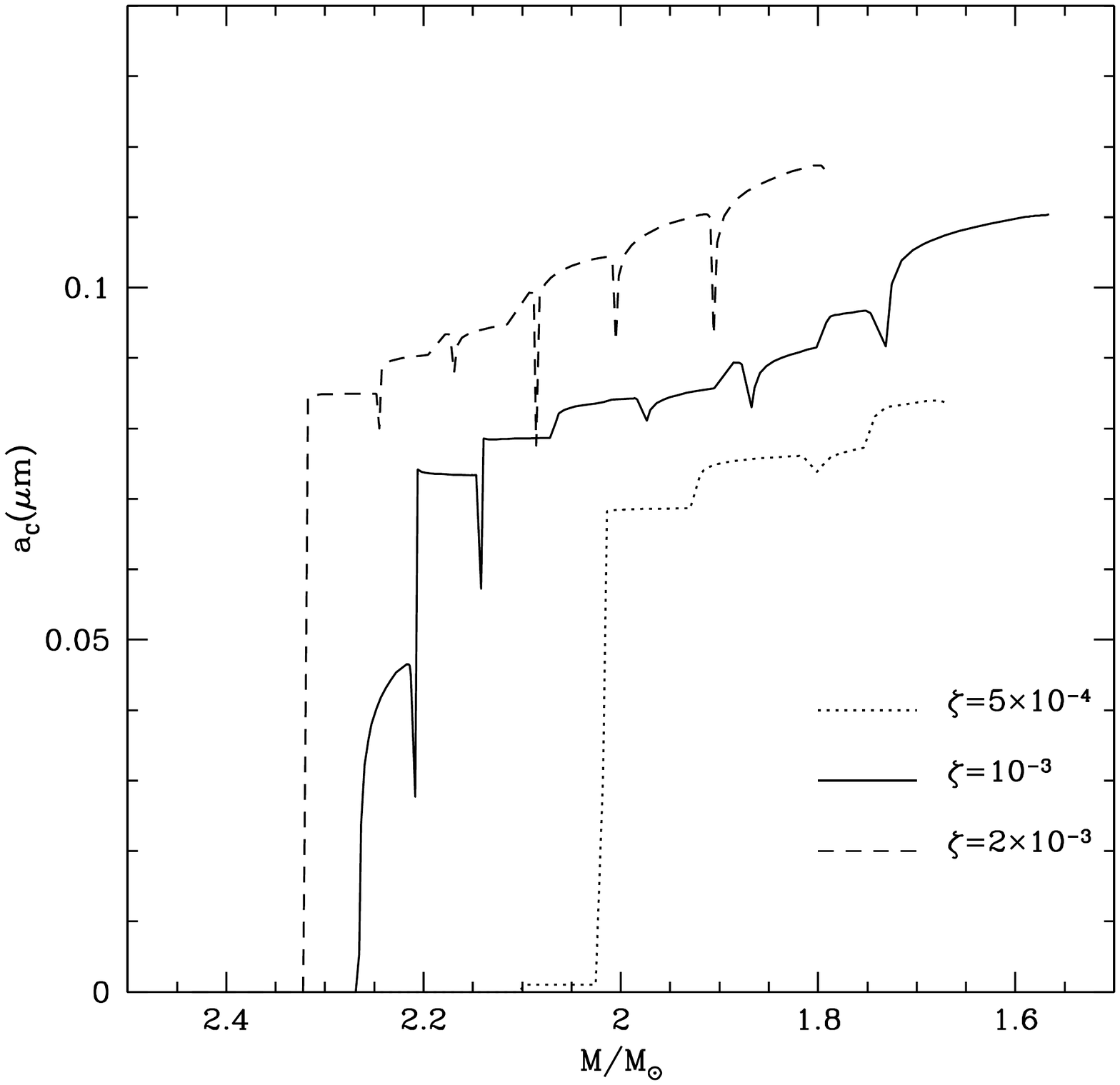}}
\end{minipage}
\begin{minipage}{0.45\textwidth}
\resizebox{1.\hsize}{!}{\includegraphics{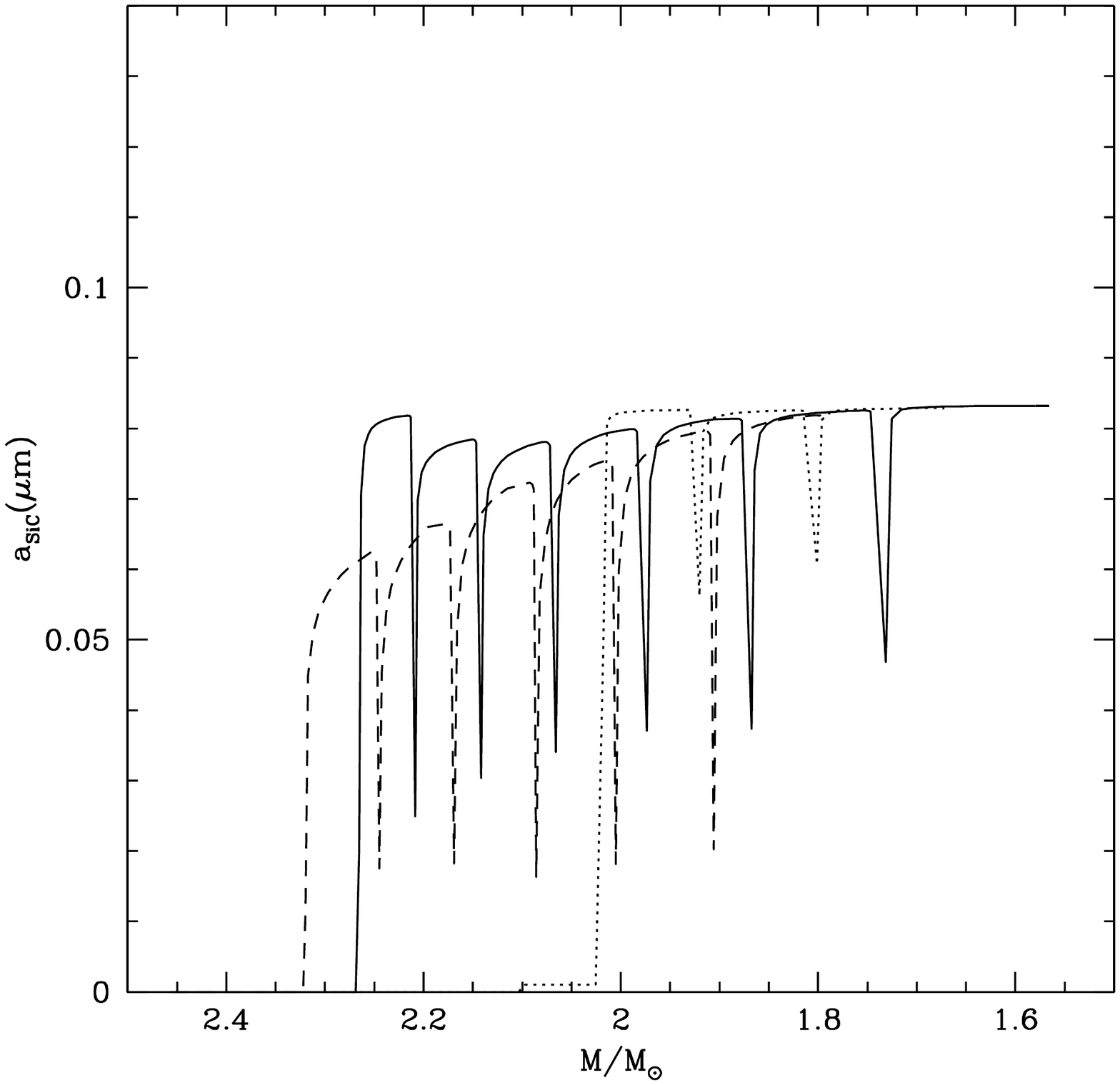}}
\end{minipage}
\vskip-40pt
\caption{The variation of the size of solid carbon (left) and SiC (right) grains
formed during the evolution of a model of initial mass 2.5M$_{\odot}$. The three
lines indicate the results corresponding to various assumptions for the extra--mixing
from the borders of the convective shell that develops following each TP.}
\label{f3dup}
\end{figure*}

Stellar models not experiencing HBB achieve only a modest production of silicates,
because the density of the key--species declines very rapidly away from the surface of the star.
This can be seen in Fig.~\ref{foliv}, where we note the small olivine grain sizes formed by the 
$\rm 3~M_{\odot}$ model.
The possibility that carbon--type dust is produced depends on the extent of the extra--mixing
from the convective shell driven by the thermal pulse. We see from Fig.~\ref{fovsh} that 
when overshooting is not considered, it is only in the $\rm 3~M_{\odot}$ model that the condition to
produce carbon dust, i.e. C/O$>1$, is met. The situation changes in the presence of even 
a modest amount of extra--mixing (i.e. $\zeta=0.001$). In such case, TDU becomes much more efficient 
and penetrating, and the stars not experiencing HBB reach the C--star stage, producing carbon--type dust.
The fraction of carbon condensed into dust, of the order of $f_C \sim 0.1$, is smaller than
predicted by more detailed models of dust--driven mass loss (see e.g. \citet{mattsson2011}).
This is partly due to the small extent of the extra--mixing assumed during the TDU,
that prevents great enhancements of the surface carbon.

\section{How robust are the present results?}
The results found in terms of the type, total mass, and grain size distribution of the
dust particles formed around AGBs depend on many assumptions made to calculate the
evolutionary sequences, associated with the details of the AGB modelling and the 
description of the dust formation process.

\subsection{The physical inputs for the AGB description}
In paper I (section 5) we explored how the results obtained depend on the assumptions in 
the macro--physics adopted to describe the AGB evolution, i.e the treatment of convection 
(both the efficiency of the convective transport and the treatment of convective borders) 
and the mass loss prescription.

The convection model is still the main source of uncertainty in the
more massive models, because it determines the extent of the HBB experienced, and thus the
surface evolution of two key--elements for the dust formation process, i.e. carbon and oxygen.
Use of a lower convection efficiency would increase the threshold mass separating the
stars predicted to form silicates from those producing carbon grains. 

For the models not undergoing HBB, the treatment of the convective borders is the key--issue
in determining how much dust is formed in their surroundings. In the present work, the possible
overshoot of the convective eddies into the regions radiatively stable is described by the
parameter $\zeta$ (see Sect. 2.1). The two panels of Fig.~\ref{f3dup} show the effects of
changing $\zeta$ on the size of the solid carbon (left panel) and SiC (right) grains. 
The results refer to a 2.5M$_{\odot}$ model, in the range of masses experiencing TDU. We see 
that an increase in $\zeta$ by a factor 2 leads to formation of carbon grains that 
are 10--20$\%$ bigger. 
In terms of the amount of dust formed, for the three models discussed we find masses of 
solid carbon (the main dust component in the present case) of $2.8\times 10^{-4}M_{\odot}$, 
$6.4\times 10^{-4}M_{\odot}$, $9\times 10^{-4}M_{\odot}$, for $\zeta=5\times 10^{-4}$, 
$10^{-3}$, $2\times 10^{-3}$, respectively. These results suggest that the extension of 
the convective shell formed at each TP has a critical impact on the amount of carbon--dust 
formed in AGBs experiencing TDU, and is the main source of uncertainty for the predictive
power of this work.

Mass loss was also shown to be a source of uncertainty in paper I, because use of a 
prescription with a milder dependence on the luminosity than the Bl\"ocker's recipe
(e.g. the formulation by \citet{VW93}) would lead to a lower production of silicates: 
this would greatly affect the resultant dust yields found, because the stars would loose only a tiny
fraction of their envelope by the time that the surface oxygen is consumed, so that most of
the mass lost would be oxygen--poor, which prevents the formation of silicates. We note that
this effect would be of smaller importance here, because the extent of HBB is softer, such that
the oxygen is not severely depleted (see the right panel of Fig.\ref{fagb}).

To investigate also how the treatment of mass loss influences the production of 
carbon--rich dust in models of smaller mass, we calculated models that
reach the C--star stage (initial masses $2$, $2.5$, $3M_{\odot}$) with the mass loss 
formulation by \citet{watcher08}. This treatment is particularly suitable for our purposes,
because it is based on the same metallicity as the models presented here (Z=0.008), and
is aimed at determining the rate of mass loss for AGB stars already in the C--star phase.

The treatment by \citet{watcher08} predicts that during the C--star phase mass is lost faster in 
comparison to the Bl\"ocker prescription, because of the large negative exponent of the
effective temperature in the \citet{watcher08} formulation (see their Eq.1). The star experiences 
a smaller number of thermal pulses, which prevents the achievement of large C/O ratios. This can 
be clearly seen in the left panel of Fig.~\ref{fovsh}, where evolutions calculated by means 
of the \citet{watcher08} formula are shown with dashed lines, to be compared with the dotted
tracks. We note in the right panel that the amount of solid carbon formed is similar in the two 
cases, because the effects of the larger abundance of surface carbon achieved by the 
Bl\"ocker's models are partly compensated by the smaller mass loss rate experienced, 
which, in agreement with our scheme, leads to a smaller production of dust.

\subsection{The optical properties of silicates}

The quantity $\Gamma$, defined in Eq.~\ref{eqgamma}, represents the coupling 
between matter and radiation field. The increase in $\Gamma$ when dust begins to form 
leads to an increase in the radiation pressure, with the consequent acceleration of 
the wind, that halts the dust formation process.

\begin{figure}
\resizebox{1.\hsize}{!}{\includegraphics{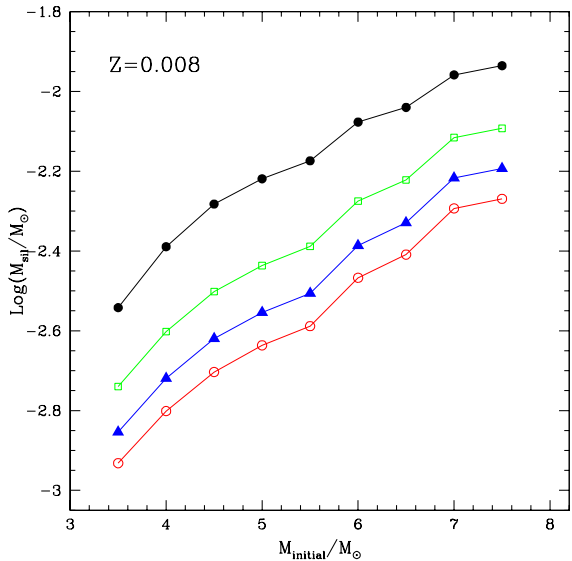}}
\vskip-40pt
\caption{The total mass of silicates produced by AGB and SAGB models experiencing HBB. The various 
symbols indicate the results obtained with different sets of optical properties:
open (red) circles \citet{jones76}; full (blue) triangles \citet{ossenkopf92};
open (green) squares \citet{draine84}; full (black) dots \citet{dorschner95} (olivine),
\citet{brewster92} (quartz), and \citet{jager94} (pyroxene).}
\label{fkappa}
\end{figure}

The interaction between the radiation and the dust grains is expressed by the coefficient
$k$, which, in turn, depends on the optical properties of the type of dust formed.
For silicate grains, the uncertainty associated to the refractive index is
still large and it is important to estimate how this affects the predicted mass of dust
formed by AGB and SAGB stars.
We therefore repeat our analysis adopting different sets of silicates refractive index
currently available in the literature, such as the optical properties of the astronomical 
silicates by \citet{draine84}, the empirically determined efficiencies by \citet{jones76},
the optical constants by \citet{ossenkopf92}, and the invidual properties for each of the three
silicates considered here, i.e. olivine \citep{dorschner95}, pyroxene \citep{jager94}, and
quartz \citep{brewster92}.
A detailed analysis of the differences among these data can be found in \citet{jeong03} 
(see their Figs.~2 and 3). Here we simply compare the results that we obtain for the total 
mass of silicates produced, restricting the comparison to stars with mass 
$\rm M \geq 3.5~M_{\odot}$ that experience HBB.

The results are shown in Fig.~\ref{fkappa}. The largest mass of silicates is predicted to form
when the data from \citet{dorschner95} are used, whereas the description by \citet{jones76} 
provides the lowest silicates production. We note that, independently of the
initial mass of the star, the difference is within $\sim 0.4$ dex; we assume this 
to be the degree of uncertainty due the choice of the refractive index.

\section{Discussion}
\begin{figure*}
\begin{minipage}{0.45\textwidth}
\resizebox{1.\hsize}{!}{\includegraphics{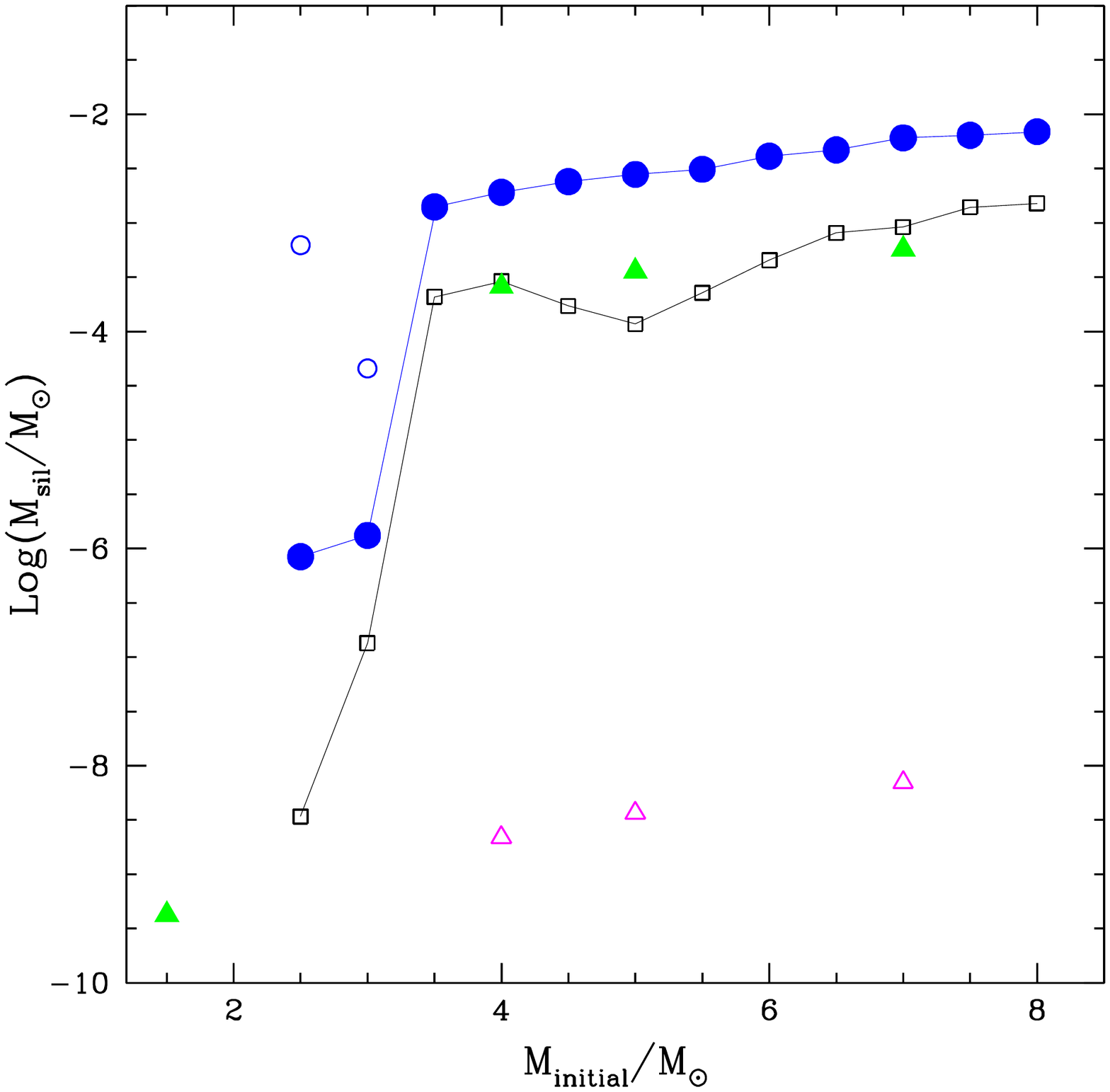}}
\end{minipage}
\begin{minipage}{0.45\textwidth}
\resizebox{1.\hsize}{!}{\includegraphics{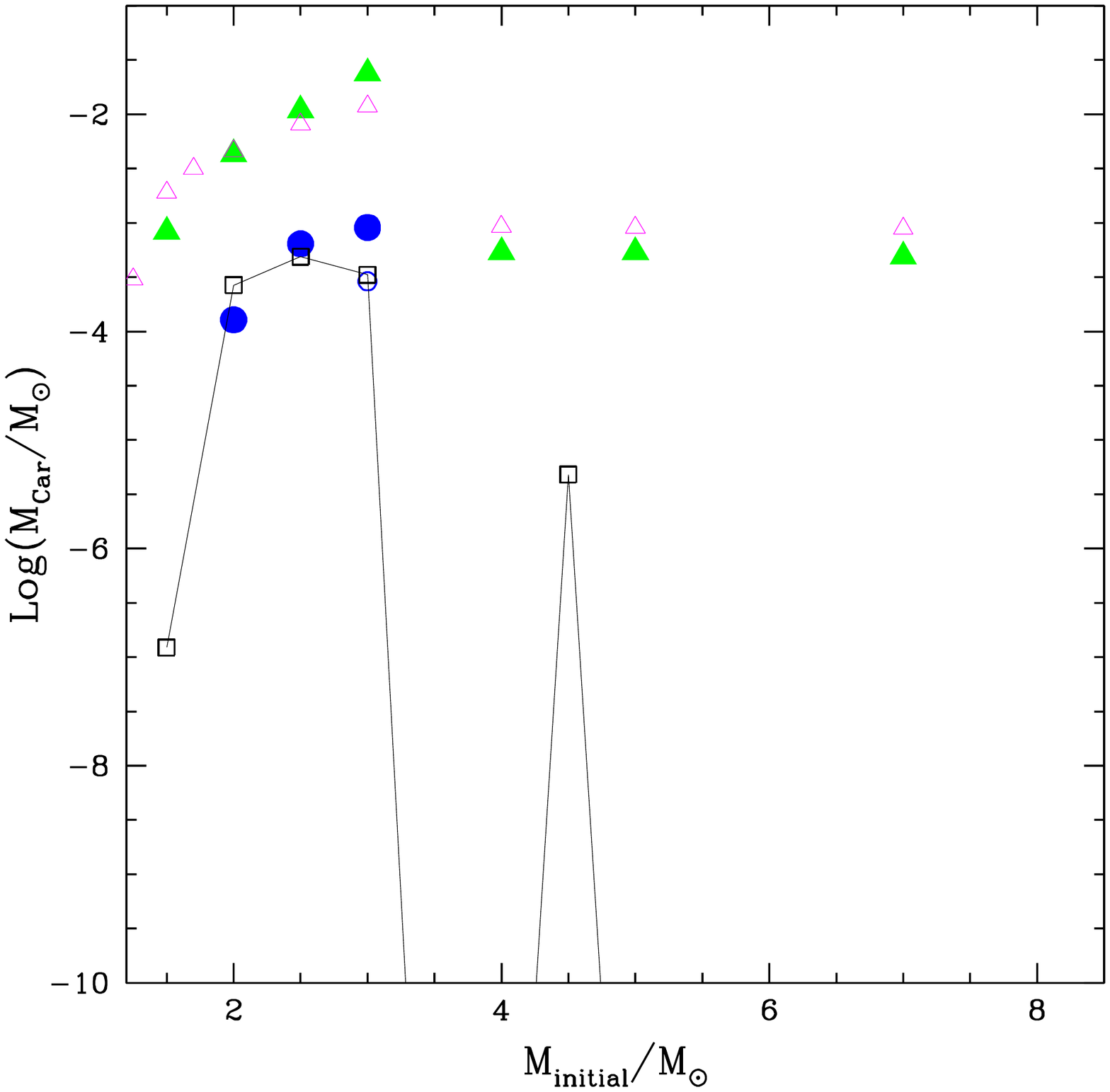}}
\end{minipage}
\vskip-40pt
\caption{Total mass of silicates ({\it left panel}) and carbon dust ({\it right panel}) 
produced by the stellar models presented in this paper (blue filled dots) and by the 
$Z=0.001$ models published in paper I (black open squares). For comparison, we also
show the results by \citet{fg06} at metallicites $Z=0.001$ (magenta open triangles) 
and $Z=0.008$ (green filled triangles). The two open points in the left panel show what
would be the mass of silicates produced by the two lowest mass models with $Z=0.008$ if
the overshoot from the convective shell were to be neglected (see the text).
}
\label{fmdust}
\end{figure*}

The predicted mass of dust produced by AGB and SAGB stars with $Z=0.008$ is reported in 
Table~\ref{dustmass}, where we also specify the mass of the individual dust species. 
For the silicates we used the set of optical constants by \citet{ossenkopf92}. 
In Fig.~\ref{fmdust}, we show separately the predicted mass of silicates (left panel) 
and carbon--rich dust (right panel) for each stellar mass. Filled circles indicate the 
mass of dust produced by the models presented in this work, while open squares represent 
the $Z=0.001$ models discussed in paper I. For comparison, we also show the dust mass 
predicted by \citet{fg06} for the same two metallicities (open triangles $Z=0.001$; 
filled triangles $Z=0.008$).

The large production of silicates in higher--mass models, found in paper I for $Z=0.001$, is
confirmed at $Z=0.008$. The two lines in the left panel of Fig.~\ref{fmdust}, corresponding to 
the two metallicities, follow similar trends, and show a drop for masses 
$\rm M < 3.5~M_{\odot}$ that do not experience any HBB. The discontinuity at $\rm 3.5~M_{\odot}$ 
would be smaller without overshoot from the shell (see the open points).
Z=0.008 models produce a mass of silicate--type dust a factor $\sim 10$ larger compared 
to the $Z=0.001$ case. This reflects the difference in metallicity, that results in a higher 
density of silicon in the wind. The trend of dust formed with stellar mass is monotonic here, 
as opposed to the $Z=0.001$ models (note the local minimum for Z=0.001 at M=5M$_{\odot}$), 
because the HBB is softer. Thus we never reach a situation where shortage of water 
molecules inhibits silicates formation.

Comparing the mass of silicates predicted by our models with the results by \citet{fg06} at the same stellar
metallicity, we note that the difference is smaller at $Z=0.008$ than at $Z=0.001$. In fact, 
the higher metallicity prevents (or delays) the achievement of the C--star stage, so that in the more massive
stars silicates are produced independently of the physical input used to model the AGB evolution. 
Still, comparing the filled circles and the filled triangles, we note that the mass of silicates predicted 
by our models is larger because ({\it i}) the HBB experienced is stronger and ({\it ii}) the production of 
silicates in the models by \citet{fg06} is limited to the first part of the evolution, before the star 
enters the C--star stage. The results by \citet{fg06} do not show the same proportionality
of the mass of silicates formed with metallicity.

\begin{table*}
\caption{Dust mass produced by AGB and SAGB models of metallicity $Z=0.008$. The initial stellar 
mass $M$ is reported in the first column. The total mass of dust, $M_{\rm d}$ and
the mass of olivine ($M_{\rm ol}$), pyroxene ($M_{\rm py}$), quartz ($M_{\rm qu}$), solid iron 
($M_{\rm ir}$), solid carbon ($M_{\rm C}$) and silicon carbide ($M_{\rm SiC}$) are also shown. All 
the masses are expressed in solar units. The optical constants from \citet{ossenkopf92} were
used for the silicates. The three sets of models differ in the mass loss treatment (we used the
recipes from \citet{blocker95} and \citet{watcher08}) and the
treatment of the convective borders (expressed by the parameter $\zeta$).}
\label{dustmass}
\begin{tabular}{cccccccc}
\hline
\hline
$M$ & $M_{\rm d}$ & $M_{\rm ol}$ & $M_{\rm py}$ & 
$M_{\rm qu}$ & $M_{\rm ir}$ & $M_{\rm C}$ & $M_{\rm SiC}$  \\
\hline
&&&& Bl\"ocker -- $\zeta=0.001$ &&&  \\
\hline
1.5  &  1.15D-05  &  4.97D-07  &  1.89D-07  &  6.97D-08  &  1.06D-05  &  1.23D-07  &  3.85D-08  \\
2.0  &  1.46D-04  &  1.01D-12  &  6.72D-13  &  3.22D-13  &  1.85D-05  &  1.28D-04  &  0.00D+00   \\
2.5  &  8.30D-04  &  5.46D-07  &  2.06D-07  &  8.95D-08  &  1.89D-06  &  6.42D-04  &  1.86D-04   \\
3.0  &  1.14D-03  &  7.54D-07  &  3.95D-07  &  1.67D-07  &  2.60D-06  &  9.06D-04  &  2.33D-04   \\
3.5  &  1.41D-03  &  1.03D-03  &  3.22D-04  &  4.94D-05  &  7.77D-05  &  0.00D+00  &  0.00D+00   \\
4.0  &  1.92D-03  &  1.44D-03  &  4.19D-04  &  4.52D-05  &  6.62D-05  &  0.00D+00  &  0.00D+00   \\
4.5  &  2.41D-03  &  1.87D-03  &  5.00D-04  &  3.86D-05  &  5.43D-05  &  0.00D+00  &  0.00D+00   \\
5.0  &  2.80D-03  &  2.20D-03  &  5.53D-04  &  3.69D-05  &  4.57D-05  &  0.00D+00  &  0.00D+00   \\
5.5  &  3.12D-03  &  2.50D-03  &  5.88D-04  &  3.64D-05  &  4.21D-05  &  0.00D+00  &  0.00D+00   \\
6.0  &  4.11D-03  &  3.47D-03  &  6.16D-04  &  1.95D-05  &  2.24D-05  &  0.00D+00  &  0.00D+00   \\
6.5  &  4.69D-03  &  4.07D-03  &  5.93D-04  &  2.36D-05  &  2.52D-05  &  0.00D+00  &  0.00D+00   \\
7.0  &  6.07D-03  &  5.49D-03  &  5.68D-04  &  1.26D-05  &  2.10D-05  &  0.00D+00  &  0.00D+00   \\
7.5  &  6.41D-03  &  5.82D-03  &  5.76D-04  &  1.36D-05  &  2.35D-05  &  0.00D+00  &  0.00D+00   \\
8.0  &  6.92D-03  &  6.11D-03  &  6.40D-04  &  1.70D-05  &  2.64D-05  &  0.00D+00  &  0.00D+00   \\
\hline
&&&& Bl\"ocker -- $\zeta=0$ &&&  \\
\hline
2.0  &  2.40D-04  &  1.99D-14  &  1.23D-14  &  4.95D-15  &  2.40D-04  &  0.00D-00  &  0.00D+00   \\
2.5  &  8.00D-04  &  4.30D-04  &  1.49D-04  &  4.52D-05  &  1.75D-04  &  0.00D-00  &  0.00D-00   \\
3.0  &  6.43D-04  &  2.90D-05  &  1.15D-05  &  5.11D-06  &  4.72D-05  &  2.89D-04  &  2.61D-04   \\
\hline
&&&& Watcher (2008) -- $\zeta=0.001$ &&&  \\
\hline
2.0  &  1.88D-04  &  2.20D-12  &  1.01D-12  &  4.81D-13  &  4.77D-05  &  1.40D-04  &  0.00D+00   \\
2.5  &  8.74D-04  &  7.90D-07  &  2.06D-07  &  8.95D-08  &  1.89D-06  &  6.73D-04  &  1.98D-04   \\
3.0  &  1.03D-03  &  1.09D-06  &  3.95D-07  &  1.67D-07  &  2.60D-06  &  7.93D-04  &  2.31D-04   \\
\hline
\end{tabular}
\end{table*}

The right panel of Fig.~\ref{fmdust} shows that for the models presented in this paper 
we do not find any carbon--type dust for $M \geq 3.5~M_{\odot}$, because the HBB destroys 
the surface carbon. This is at odds with the results by \citet{fg06}, where some carbon--type dust 
is produced at all stellar masses. For carbon--type dust, the mass predicted by our models show a 
negligible dependence on the initial stellar metallicity. In fact, dust production is mainly 
sensitive to the amount of carbon formed in the $3\alpha$ burning shell and dredged--up to the surface; 
such nucleosynthesis is scarcely dependent on the initial metal content of the star, unlike the CNO burning 
at the bottom of the envelope, which is sensitive to the initial abundances of silicon, magnesium and oxygen. 

The models by \citet{fg06} predict a larger mass of carbon--type dust for low--mass stars, 
because of the different assumptions concerning the extent of the TDU. Part of the 
difference is also due to the low--T surface opacities used by the synthetic models upon 
which the AGB evolution by \citet{fg06} is computed: they are based on a solar-scaled 
mixture, where any variation in the C/O ratio is neglected. As shown by \citet{marigo02}, 
when C/O exceeds unity, CN and C$_2$ molecules replace water as the main absorbers of 
radiation, which, in turn, leads to an increase in the opacity for regions with 
temperatures below 3000K. Because a higher opacity favours cooling and expansion of
the external layers, this will reflect into an increase in the rate at which
mass loss occurs. In our models we therefore expect that less carbon is
formed, because the star looses its envelope more rapidly.

\begin{figure}
\resizebox{1.\hsize}{!}{\includegraphics{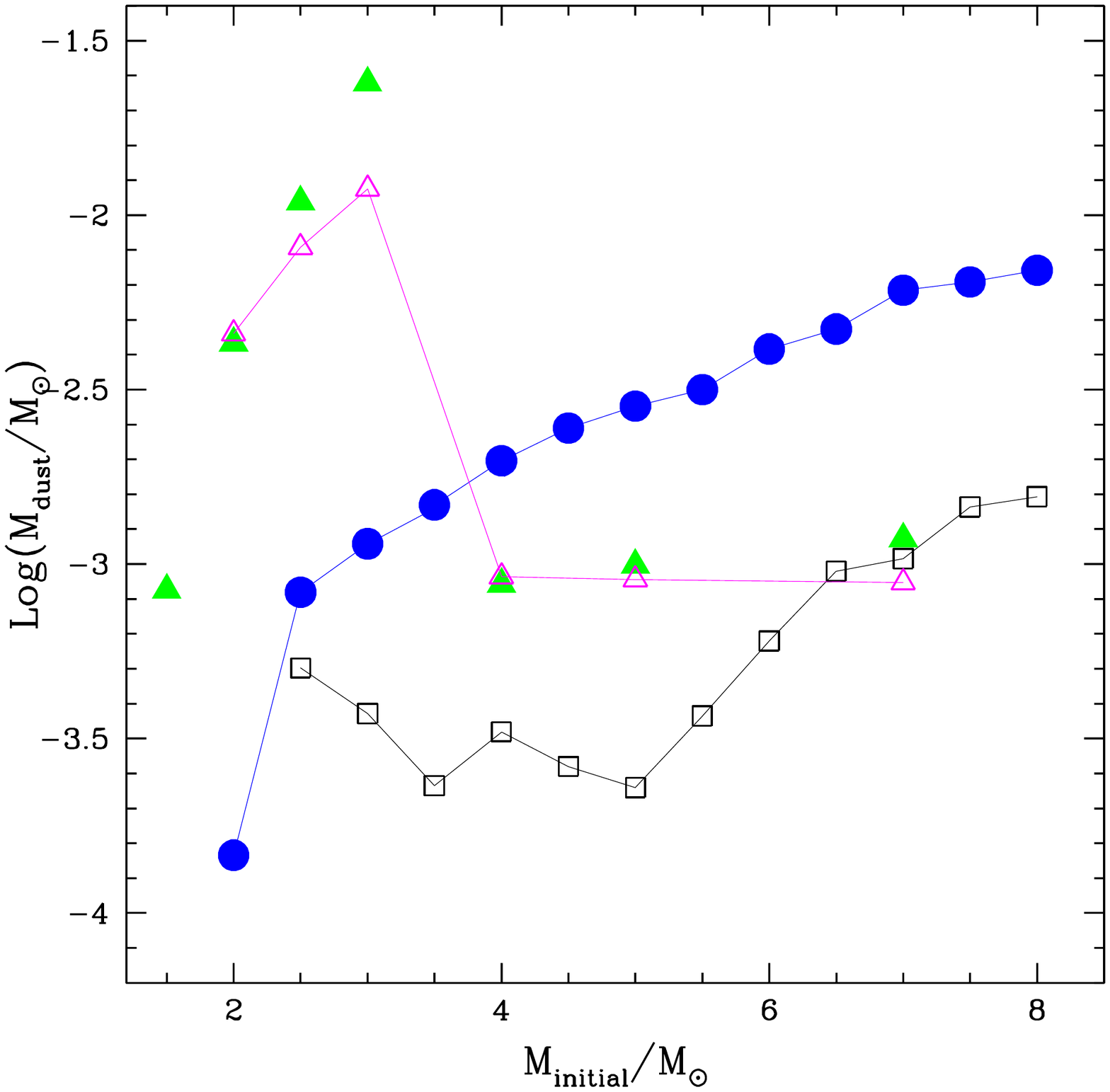}}
\vskip-40pt
\caption{The total mass of dust produced by stars with different initial masses and metallicities. 
The meaning of the various symbols is the same as in Fig.~\ref{fmdust}.
}
\label{fdusttot}
\end{figure}

Fig.~\ref{fdusttot} shows the total mass of dust produced by stars with different initial mass and
metallicity. Compared to \citet{fg06}, our models show an opposite trend with stellar mass, which is 
both qualitative and quantitative: more massive models produce more dust, under the form of silicate--type
grains, whereas in the \citet{fg06} case most of the dust is produced by low--mass stars, and
as carbon dust. In addition, our models appear to be extremely sensitive to the initial stellar 
metallicity, with higher $Z$ models producing more dust. On the contrary, the mass of dust predicted
by \citet{fg06} is relatively constant with $Z$, because the surface--carbon abundance in those models
is mostly dependent on the nucleosynthesis in the shell that forms during the thermal pulse.

\subsection{Implications for the cosmic dust yields}

\begin{figure}
\resizebox{1.\hsize}{!}{\includegraphics{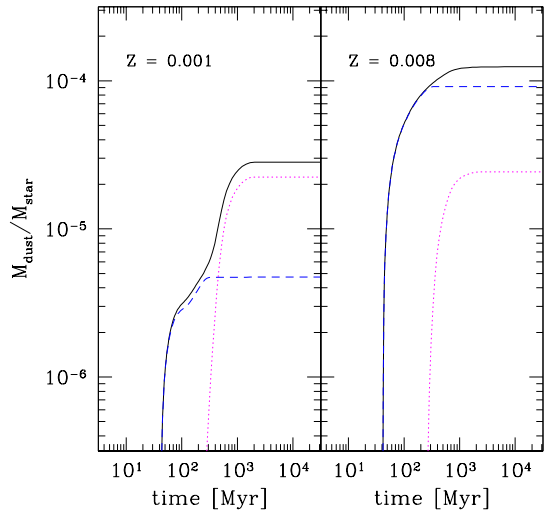}}
\vskip-40pt
\caption{The total mass of dust produced by AGB and SAGB stars normalized to the total mass of stars
formed in a single burst at $\rm time = 0$ with a Larson IMF ranging between $1$ and $100 \rm M_{\odot}$
and initial metallicity of $Z=0.001$ ({\it left panel}) and of $Z=0.008$ ({\it right panel}). 
The solid lines indicate the total mass of dust, the dotted and dashed lines show the separate contributions
of carbon and silicate dust. The stellar evolutionary timescales are computed from the ATON stellar evolutionary
model.
}
\label{fig:cosmicyield}
\end{figure}

Following the analysis done in paper I, we can investigate the implications of the above findings for the
cosmic dust enrichment contributed by intermediate mass stars. To do this, we assume that all the stars 
form in a single burst at a reference initial time with metallicity $Z=0.008$ and according to 
a Larson initial mass function (IMF) with stellar mass in the range $\rm [0.1 - 100]~M_{\odot}$ 
and a characteristic stellar mass of 
$m_{\rm ch}=0.35 \rm M_{\odot}$  (see equations 20 and 21 in paper I and Valiante et al. 2009).
The cosmic dust yield of stars with masses $M \le 8 M_{\odot}$, i.e. 
the total dust mass injected into the interstellar medium by AGB and SAGB stars 
normalized to the total mass of stars formed in the burst, evolves
on stellar evolutionary timescales (see Fig.~\ref{fig:cosmicyield}, right panel).
The dotted and dashed lines indicate the separate contributions of carbon and silicate dust. 
For comparison, we show in the left panel the same quantities obtained for $Z=0.001$ stars\footnote{The
values reported are the same as those shown in Fig. 13 of paper I, where the small extra contribution of silicate dust
production at late evolutionary timescales (that is not apparent in Fig.\ref{fig:cosmicyield}) is due to a numerical
error.}.

At the end of their evolution, intermediate mass stars with higher metallicity produce a cosmic dust yield that 
is a factor $\sim 4.4$ larger than that associated with their lower metallicity counterparts. In addition, when
$Z=0.008$ stars are considered, the dust mass is dominated by silicates at all times. The small carbon dust 
yields produced by stars with masses $\le 3 \rm M_{\odot}$ is not compensated by the higher frequency 
of low-mass stars obtained with the adopted stellar IMF. Conversely, the large silicate dust yields produced
by more massive stars leads to a prompt enrichment on timescales $\ge 40$~Myr and, at $\sim 100$~Myr 
after the burst, the mass of silicate dust produced by $Z=0.008$ stars is more than a factor 
10 larger than that released by $Z=0.001$ stars.

Our findings may have important implications for chemical evolution models with dust. In particular, 
it was suggested that polycyclic aromatic hydrocarbons (PAHs) and carbon dust are mostly produced in 
AGB stars \citep{dwek98}. This has offered an explanation for the observed correlation of PAH 
line intensities with metallicity in nearby galaxies \citep{madden06}. The correlation has been
interpreted as a trend of PAH abundance with galactic age, reflecting the delayed injection of PAHs and carbon dust 
into the ISM by AGB stars in their ﬁnal, post-AGB phase of their evolution \citep{galliano08}.
Our results seem to suggest that the contribution AGB and SAGB to carbon dust enrichment may be significantly 
less important than previously estimated. The observed correlation may be best attributed to 
the destruction of PAH molecules by photoevaporation or photodissociation, that are more efficient in 
low-metallicity environments. A detailed comparison between the yields predicted by the present models and
those predicted for massive stars that explode as supernovae \citep{bianchi07} is deferred to a future study.

\section{Conclusions}
We have calculated the dust formation around intermediate mass stars of metallicity Z=0.008 in 
the mass range $1M_{\odot} \leq M \leq 8M_{\odot}$ during their whole AGB (or SAGB) phase.

We confirm the main finding of a previous exploration based on a smaller metallicity, i.e. 
that more massive objects, experiencing HBB, achieve a rich production of
silicates, favoured by the strong mass loss experienced. Lower--mass stars, on the
other hand, after an early phase with a minor production of silicates, will be
surrounded by carbon grains, once the surface C/O ratio exceeds unity, as a 
consequence of repeated TDU episodes.

The amount of silicates produced depends on the metallicity of the stars: compared to
the Z=0.001 models analyzed in our previous paper, we expect a much larger dust  
production here, due both to the larger silicon mass fraction, and to the softer HBB
experienced, that prevents total destruction of the water molecules present in the
wind, required to produce any kind of silicate.

The major sources of uncertainty in the amount of silicates produced are the
treatment of convection, with the relative strength of HBB, and the poor knowledge
of the optical constants of the various silicates formed, which reflect into an
uncertainty of $\sim 0.4$dex. On the other hand, under the C--star regime, the most 
important source
of uncertainty in the quantity of carbon dust formed is the extent of the TDU,
which is poorly known from first principles: a tiny amount of extra--mixing from the
borders of the convective shell developed during the TPs favours a much larger inwards
penetration of the convective envelope, and leads to much stronger TDUs; the quantity
of dust produced changes dramatically, which confirms the poor robustness of the results
obtained in this range of masses.

We also like to point out that the assumed mass-loss rate is (in the present framework) 
important for the resultant degree of dust condensation as well as the time scale on which the 
stellar envelope is lost from the star. That, in turn, may affect the dust yield we obtain.

Based on these results, we find that the cosmic dust yield at this metallicity is
higher by a factor $\sim 5$ compared to the Z=0.001 case discussed in our previous
work, and is dominated by silicates at all times. Silicates dominate the cosmic dust yield 
because of the observed monotonic trend with stellar mass of the total dust mass produced by 
AGB and SAGB stars.  In this scenario, the AGB/SAGB dust production (in which silicates are the 
favored species produced) dominates over presence of carbon dust grains, the favoured type 
of dust to be produced by lower-mass stars.  This is true even with the larger number of 
lower-mass stars expected according to any realistic IMF.

\section*{Acknowledgments}
The authors are indebted to Paola Marigo, for the computation of the low--temperature
opacities in the C--star regime, by means of the AESOPUS tool. We thank the anonymous
referee for the careful reading of the manuscript, and the many comments and suggestions,
that improved the quality of the paper. MDC acknowledges financial support from the 
Observatory of Rome.

\end{document}